\documentclass{mem}
\usepackage{natbib}\usepackage{txfonts}\usepackage{balance}
\usepackage{graphicx}
\idline{75}{282}
\begin{document}
\def\teff{$T\rm_{eff }$}
\def\kms{$\mathrm {km s}^{-1}$}

\title{The ~mass-­metallicity ~relation ~of  ~galaxies ~up ~to ~redshift ~0.35}

\author{Ivo ~Saviane\inst{1} \and Irina ~Yegorova\inst{1} \and Dominique ~Proust\inst{2}
\and Fabio ~Bresolin\inst{3} \and Valentin ~Ivanov\inst{1} \and
Enrico ~V. ~Held\inst{4} \and John ~Salzer\inst{5} \and R. ~Michael ~Rich\inst{6}}

\offprints{I. Saviane }

\institute{European Southern Observatory, Alonso de Cordova 3107, Santiago, Chile\\
\email{isaviane@eso.org} 
\and 
Observatoire de Paris-Meudon, GEPI, F92195 MEUDON, France
\and Institute for Astronomy, 2680 Woodlawn Drive, Honolulu, HI 96822, USA
\and Astronomical Observatory, vicolo Osservatorio 5, Padova, Italy 
\and Department of Astronomy, Indiana University, 727 East Third Street, Bloomington, IN
47405, USA
\and Division of Astronomy \& Astrophysics, University of California, 430 Portola Plaza, 
Los Angeles, CA 90095-1547, USA}

\authorrunning{I. Saviane}

\titlerunning{MZR up to $z=0.35$}

\abstract{Our research on the age-metallicity and mass-metallicity
  relations of galaxies is presented and compared to the most recent
  investigations in the field. We have been able to measure oxygen
  abundances using the direct method for objects spanning four orders of
  magnitude in mass, and probing the last 4 Gyr of galaxy evolution.  We
  have found preliminary evidence that the metallicity evolution is
  consistent with expectations based on age-metallicity relations
  obtained with low resolution stellar spectra of resolved Local Group
  galaxies.

\keywords{Galaxies: abundances --
Galaxies: evolution --
Galaxies: fundamental parameters --
Galaxies: ISM --
Galaxies: star formation
}
}
\maketitle{}

\section{Introduction}

Forty  years ago \citet{Larson1974} attempted one of the first theoretical
investigations to explain the mass-metallicity relation (MZR) of (elliptical)
galaxies.  His final relation is remarkably similar to current observational
results: the metallicity goes from {[}m/H{]}$\sim-2.3$ at $M=10^{5}M_{\odot}$
to {[}m/H{]}$\sim0$ at $M=10^{13}M_{\odot}$, and it flattens at
high masses. To test these kind of predictions we can either reconstruct
the age-metallicity relation (AMR) of galaxies of different masses,
or go back in time by taking snapshots of the MZR at different redshifts.

\section{Age-metallicity relations of resolved galaxies\label{sec:AMRs-of-resolved}}

\begin{figure*}[t]
\begin{centering}
\includegraphics[width=1\textwidth]{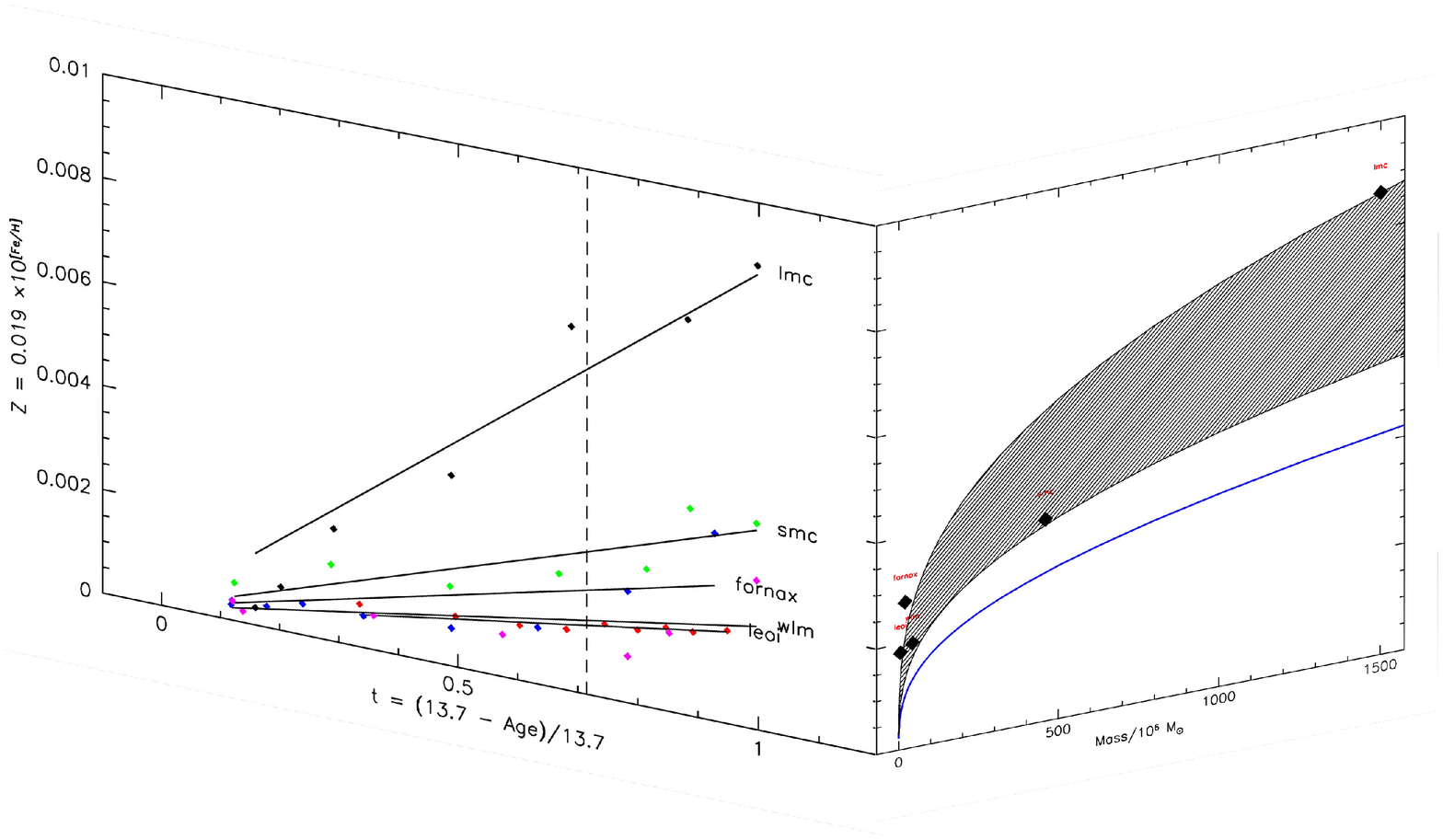}
\par\end{centering}

\caption{The left side of the box shows age-metallicity relations from \citet{Leaman2013_TheComparativeChemicalEvolutionofanIsolatedDwarfGalaxy:AVLTandKeckSpectroscopicSurveyofWLM}
and \citet{Gullieuszik2009}. The vertical dotted line marks redshift
$z=0.35$. The right side shows the terminal points of the linear
AMRs against present galaxy mass (with mass from \citealt{McConnachie2012_TheObservedPropertiesofDwarfGalaxiesinandaroundtheLocalGroup}).
The blue curve is the mass-metallicity relation from \citet{Kirby2011_Multi-elementAbundanceMeasurementsfromMedium-resolutionSpectra.III.MetallicityDistributionsofMilkyWayDwarfSatelliteGalaxies},
which represents the evolutionary status of galaxies some time in
the past. The shaded area shows the effect of evolution assuming that
current metallicities are a factor $1.6$ higher than those of K11
(the area encompasses the $\pm1\sigma$ error from K11). \label{fig:AMR-and-MZR}}
\end{figure*}

Thanks to the advent of moderate resolution, multiplexing spectrographs
at 10m-class telescopes, AMRs are now available for several resolved
galaxies in the Local Group. As an example, in \citet{Gullieuszik2009}
we measured the metallicity of 54 red-giant stars in the dSph galaxy
Leo~I by applying the CaT method (\citealp{Armandroff1991a}), and
we found individual ages by interpolating a set of isochrones. Similar
methods are employed by other groups: for example \citet{Leaman2013_TheComparativeChemicalEvolutionofanIsolatedDwarfGalaxy:AVLTandKeckSpectroscopicSurveyofWLM}
published AMRs for Fornax, WLM, and the Magellanic clouds. These relations
can be seen in Fig.~\ref{fig:AMR-and-MZR} together with that of
Leo~I. The figure shows that a linear growth of $Z$ vs. time is
a reasonable approximation, and in addition, if the end points of
the AMRs are plotted vs. mass, then there is a good correlation between
mass and metallicity. Therefore as long as stellar mass grows with
time, one can expect a well defined MZR at any age (redshift), at
least as far as dwarf galaxies are concerned. In particular at redshift
$z=0.35$ the universe is $\sim70\%$ its current age so we can expect
a factor 1.4 increase in $Z$ since that time, or 0.14 dex in {[}m/H{]}.

\section{Snapshots of the MZR at different redshifts}

\begin{figure*}
\begin{centering}
\includegraphics[width=0.6\textwidth]{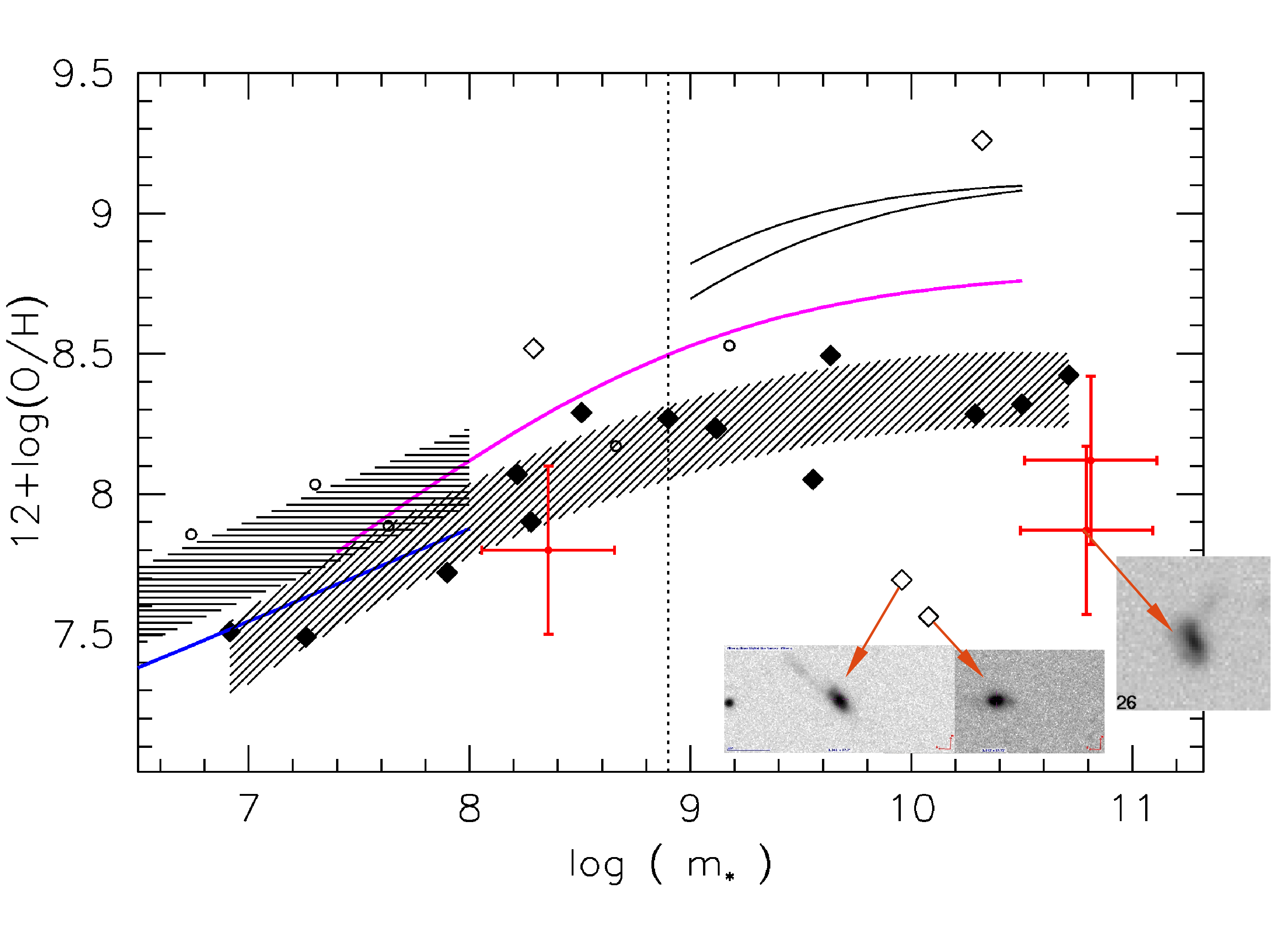}
\par\end{centering}

\caption{Several MZR are represented in this figure: the open and filled diamonds
are data from \citet{Saviane2008} (to the left of the vertical dotted
line) and from papers in preparation. The blue curve and horizontally
hashed area are the MZR from \citet{Kirby2011_Multi-elementAbundanceMeasurementsfromMedium-resolutionSpectra.III.MetallicityDistributionsofMilkyWayDwarfSatelliteGalaxies}
and the same MZR with an offset by $0.22$~dex (like in Fig.~\ref{fig:AMR-and-MZR});
the other hashed area is the MZR from Saviane et al. (in preparation).
Open small circles are present-day metallicities reached by the AMRs
of Fig. \ref{fig:AMR-and-MZR}. The magenta curve is the MZR from
\citet{Andrews2013_TheMass-MetallicityRelationwiththeDirectMethodonStackedSpectraofSDSSGalaxies},
and the two short curves are MZRs from \citet{Zahid2013_TheChemicalEvolutionofStar-formingGalaxiesovertheLast11BillionYears}
for $z=0.08$ and $z=0.29$ (upper and lower curve, respectively).
Finally red diamonds with error bars are the three star-forming galaxies
in AC114 where we could compute abundances via the direct method.
The inset images show that AC114 and KISS galaxies with abundances
that are more than $3\sigma$ lower than the average, show signs of
being accreting fresh gas. \label{fig:Several-MZR-are}}
\end{figure*}

Exploring the MZR at different redshifts is an active research field
where we are carrying out our own investigation since a few years.
Some of the major recent studies can be found in, e.g, \citet{Tremonti2004},
\citet{Pettini2004_[OIII]/[NII]asanabundanceindicatorathighredshift},
\citet{Erb2006b}, \citet{Kobulnicky2004_Metallicitiesof0.3ltzlt1.0GalaxiesintheGOODS-NorthField},
\citet{Savaglio2005}, \citet{Lee2006}, \citet{Maiolino2008}, \citet{Zahid2013_TheChemicalEvolutionofStar-formingGalaxiesovertheLast11BillionYears}.

\subsection{Local universe}

Our relation for dIrr galaxies in nearby groups has been published
in \citet{Saviane2008} (hereafter, S08), and we are currently working
on data for galaxies in the local universe, and for members of the
AC~114 cluster at redshift 0.35. The aim is to extend the MZR from
dwarf to massive galaxies, and to observe its evolution with redshift.
The main feature of our investigation is that we are trying to use
the direct method for the whole mass range, which is quite demanding
in terms of observing time because it requires to measure faint auroral
lines. This task becomes especially difficult for massive galaxies,
which are usually more metal-rich than dwarfs: when metals become
more abundant, gas cooling by atomic lines becomes more effective,
and the dropping temperature reduces emissivities. Therefore the so-called
empirical method is much more popular, because it uses strong nebular
lines that are much easier to measure. Very large galaxy samples can
thus be easily built. An interesting exception to this trend is the
recent study of \citet[hereafter AM13]{Andrews2013_TheMass-MetallicityRelationwiththeDirectMethodonStackedSpectraofSDSSGalaxies},
where SDSS spectra were stacked to boost the signal-to-noise ratio
of {[}\ion{O}{iii}{]}$\lambda$4363. In this way, the electron
temperature could be calculated even for galaxies at the top end of
the mass function, using the line ratio {[}\ion{O}{iii}{]}$\lambda$4363/{[}\ion{O}{iii}{]}$\lambda\lambda$4959,5007.

To extend our study to massive galaxies, we obtained Keck/LRIS spectra
of objects from the KISS survey (\citealp{Salzer2000}). We have been
able to measure central oxygen abundances for eleven targets, using
either the quoted oxygen line ratio, or {[}\ion{N}{ii}{]}$\lambda$5755/{[}\ion{N}{ii}{]}$\lambda\lambda$6548,6583.
Fig.~\ref{fig:Several-MZR-are} shows that these objects extend the
S08 MZR to larger masses. There are also two low-metallicity objects,
which might be accreting fresh gas that could be diluting their abundances:
indeed broad-band imaging shows clear signs of filaments departing
from the central bodies. Another feature of our project is the usage
of H-band imaging to compute target luminosities: because the SEDs
of low-mass stars peak near that band, H-band luminosities are much
less sensitive to individual star-formation histories, when compared
to optical luminosities \citep{Bell2001}. 

Figure \ref{fig:Several-MZR-are} shows two other MZRs for the local
universe, from AM13 and \citet[hereafter Z13]{Zahid2013_TheChemicalEvolutionofStar-formingGalaxiesovertheLast11BillionYears}.
The relation from Z13 is fairly representative of MZRs obtained with
the empirical method, and it is well above our own. To a lesser extent
this is true for the AM13 relation as well. An independent metallicity
constraint for massive galaxies can be given by stellar abundances
for young stars in our own Milky Way.  For example based on Cepheid
data \citet{Luck2011_TheDistributionoftheElementsintheGalacticDisk.II.AzimuthalandRadialVariationinAbundancesfromCepheids}
find a well-defined metallicity gradient reaching $\sim0.3$~dex
above solar at their minimum radius of $\sim4$~kpc. Extrapolating
their linear relation of $-0.055$~dex~kpc$^{-1}$ one might expect
{[}Fe/H{]}=0.5~dex in the center. When measuring unresolved galaxies,
the observed metallicity will be averaged within the aperture of the
spectrograph: taking into account this effect, the effective gradient
becomes shallower, such that within an aperture encompassing 16~kpc
of radius a metallicity {[}Fe/H{]}=0.1~dex would be observed. Therefore,
as long as the MW represents typical massive disk galaxies, we can
expect oxygen abundances varying between $12+\log({\rm O/H})\sim8.8$
and $\sim9.2$ depending on the disk fraction that falls into the
spectrograph aperture. This range is almost the same that is comprised
between the MZRs of AM13 and Z13 at $\log M_{*}=10.5$, so based on
this exercise we would also expect the ``true'' MZR to fall between
those two relations (both based on SDSS7 data). Uncertainties of the
empirical method have been discussed, e.g., in \citet{Kewley2008_MetallicityCalibrationsandtheMass-MetallicityRelationforStar-formingGalaxies},
but it is also possible that the direct method underestimates abundances
(\citealt{Nicholls2013_MeasuringNebularTemperatures:TheEffectofNewCollisionStrengthswithEquilibriumand$kappa$-distributedElectronEnergies}).
Because our MZR stays below that of AM13 and both are based on the
direct method, it is possible that our targets are intrinsically different
than the average object of the SDSS: for example a $0.2$~dex lower
metallicity could be the attribute of objects with SFR $\sim100$
times larger than the SDSS average (\citealt{Mannucci2010_Afundamentalrelationbetweenmassstarformationrateandmetallicityinlocalandhigh-redshiftgalaxies}).
The fact that different classes of galaxies follow different MZRs
is supported by noting in Fig.~\ref{fig:Several-MZR-are} how dIrr
galaxies stay below the end points of the AMRs of dSph/dE galaxies.
dIrr galaxies evolve more slowly than dSph/dE because of their low
SFRs (e.g., \citealt{Momany2005}).

\subsection{Emission line galaxies in AC114}

Studies of the MZR are now available up to $z=3.5$ \citet{Laskar2011_ExploringtheGalaxyMass-metallicityRelationatz3-5},
but one possible problem when comparing relations at different redshifts,
is that metallicities can be obtained with different calibrations
of the empirical method, thus rendering both absolute values and relative
trends uncertain. In addition, it is not clear whether calibrations
based on local \ion{H}{ii} regions should be valid at high redshift,
where the physical conditions of the ISM might be different. To bypass
this problem we are trying to obtain direct metallicities of emission-line
galaxies in cluster AC114 at $z=0.35$. At this redshift all important
emission lines still fall in the optical range, so though $0.35$
sounds modest compared to other works, abundances can be obtained
with higher precision thus potentially revealing even a small evolution.
The 0.14~dex predicted in Sec.~\ref{sec:AMRs-of-resolved} should
certainly be within reach. Optical spectra of 153 galaxies in the
field of AC114 were obtained with 5h exposures in each of the HR~red
and MR grisms of VIMOS at the VLT. With a kinematic analysis of recession
velocities (Proust et al., in preparation) we could select 21 emission-line
galaxies over a total of 86 cluster members, and for three of them
{[}\ion{O}{iii}{]}$\lambda$4363 could be detected. We are in
the process of refining the extraction to arrive at detection in more
objects. The red crosses in Fig.~\ref{fig:Several-MZR-are} show
the location of the three galaxies in the MZ plane. At $z=0.35$ rest-frame
H-band moves into K-band, but unfortunately there is no comprehensive
NIR photometric catalog for AC114. The brightest two galaxies have
R-band COSMOS photometry (\citealp{Capak2007_TheFirstReleaseCOSMOSOpticalandNear-IRDataandCatalog};
\citealp{Tasca2009_ThezCOSMOSredshiftsurvey:theroleofenvironmentandstellarmassinshapingtheriseofthemorphology-densityrelationfromz1}),
which corresponds to V-band in rest-frame. This means that luminosities
can be affected by recent star-formation, so our masses computed with
a fiducial $M/L=2$ could be underestimated. Therefore we attributed
error~bars of a factor of two to the masses of these galaxies. Absolute
luminosities were calculated assuming a luminosity distance of 1540~Mpc.
Interestingly, one of the two luminous objects has lower abundance
than the other, but it seems to be accreting material, so it might
be another case of fresh gas diluting the ISM abundance (see Fig.~\ref{fig:Several-MZR-are}).
If the other galaxy can be considered a typical undisturbed cluster
member, then it has an oxygen abundance $\sim0.2$~dex lower than
our local MZR, roughly in agreement with the predictions of Sec.~\ref{sec:AMRs-of-resolved}.
There is no COSMOS photometry for the third galaxy, so we just assigned
it a value $R=25$, which is the faintest value of the catalog.

\section{Conclusion}

Thanks to our original dataset, and the homogeneous method of analysis,
we have been able to detect an evolution with redshift of the MZR
for the massive galaxies in our samples. If their AMRs were linear
with time, then the evolution would be consistent with the 0.14~dex
expected from reconstructed relations for less massive galaxies over
the last 4~Gyr.  Considering all possible scenarios, the true MZR
for normal disk galaxies is probably close to that of AM13. 

Fig.~\ref{fig:AMR-and-MZR} shows that the MZR is a direct consequence
of the rate of metal production $dZ/dt$ increasing with mass. In
a closed-box scenario coupled with the \citet{Schmidt1959_TheRateofStarFormation.}
law, this would not be surprising.  The model predicts that the metal
production rate is $dZ/dt=-p/M_{{\rm gas}}\, dM_{{\rm gas}}/dt=-p/M_{{\rm gas}}\,{\rm SFR}$,
where $p$ is the metal yield and $\mu=M_{{\rm gas}}/M_{{\rm TOT}}$
is the mass fraction of gas \citep{Searle1972}; and the Schmidt law
tells us that SFR is proportional to gas density. Given that gas density
is proportional to $M\, R^{-3}$, and that mass varies much more than
size among galaxies, it is expected that metal production ultimately
depends on galaxy mass. Of course gas exchanges with the environment
will happen during the life of a galaxy, so metallicity can drop below,
or can be driven above that predicted in this simple scenario. Still,
Fig.~\ref{sec:AMRs-of-resolved} seems to suggest that other processes
play a secondary role in the definition of the AMR. Note also that
Rodrigues et al. (this volume) find that intermediate mass galaxies
evolved as closed systems in the past 6~Gyr. If $Z$ increases linearly
with time, then setting $dZ/dt={\rm const}$ gives  both $M_{{\rm gas}}$
and SFR=$dM_{{\rm gas}}/dt\propto e^{-kt}$. An exponential decay
in time of the SFR could then be the symptom of a system with little
gas exchanges with the environment. 

 

 

\bibliographystyle{aa}

\begin{thebibliography}{28}
\expandafter\ifx\csname natexlab\endcsname\relax\def\natexlab#1{#1}\fi

\bibitem[{{Andrews} \&
  {Martini}(2013)}]{Andrews2013_TheMass-MetallicityRelationwiththeDirectMethod%
onStackedSpectraofSDSSGalaxies}
{Andrews}, B.~H. \& {Martini}, P. 2013, \apj, 765, 140

\bibitem[{{Armandroff} \& {Da Costa}(1991)}]{Armandroff1991a}
{Armandroff}, T.~E. \& {Da Costa}, G.~S. 1991, \aj, 101, 1329

\bibitem[{{Bell} \& {de Jong}(2001)}]{Bell2001}
{Bell}, E.~F. \& {de Jong}, R.~S. 2001, \apj, 550, 212

\bibitem[{{Capak} {et~al.}(2007){Capak}, {Aussel}, {Ajiki}, {McCracken},
  {Mobasher}, {Scoville}, {Shopbell}, {Taniguchi}, {Thompson}, {Tribiano},
  {Sasaki}, {Blain}, {Brusa}, {Carilli}, {Comastri}, {Carollo}, {Cassata},
  {Colbert}, {Ellis}, {Elvis}, {Giavalisco}, {Green}, {Guzzo}, {Hasinger},
  {Ilbert}, {Impey}, {Jahnke}, {Kartaltepe}, {Kneib}, {Koda}, {Koekemoer},
  {Komiyama}, {Leauthaud}, {Le Fevre}, {Lilly}, {Liu}, {Massey}, {Miyazaki},
  {Murayama}, {Nagao}, {Peacock}, {Pickles}, {Porciani}, {Renzini}, {Rhodes},
  {Rich}, {Salvato}, {Sanders}, {Scarlata}, {Schiminovich}, {Schinnerer},
  {Scodeggio}, {Sheth}, {Shioya}, {Tasca}, {Taylor}, {Yan}, \&
  {Zamorani}}]{Capak2007_TheFirstReleaseCOSMOSOpticalandNear-IRDataandCatalog}
{Capak}, P., {Aussel}, H., {Ajiki}, M., {et~al.} 2007, \apjs, 172, 99

\bibitem[{{Erb} {et~al.}(2006){Erb}, {Shapley}, {Pettini}, {Steidel}, {Reddy},
  \& {Adelberger}}]{Erb2006b}
{Erb}, D.~K., {Shapley}, A.~E., {Pettini}, M., {et~al.} 2006, \apj, 644, 813

\bibitem[{{Gullieuszik} {et~al.}(2009){Gullieuszik}, {Held}, {Saviane}, \&
  {Rizzi}}]{Gullieuszik2009}
{Gullieuszik}, M., {Held}, E.~V., {Saviane}, I., \& {Rizzi}, L. 2009, \aap,
  500, 735

\bibitem[{{Kewley} \&
  {Ellison}(2008)}]{Kewley2008_MetallicityCalibrationsandtheMass-MetallicityRe%
lationforStar-formingGalaxies}
{Kewley}, L.~J. \& {Ellison}, S.~L. 2008, \apj, 681, 1183

\bibitem[{{Kirby} {et~al.}(2011){Kirby}, {Lanfranchi}, {Simon}, {Cohen}, \&
  {Guhathakurta}}]{Kirby2011_Multi-elementAbundanceMeasurementsfromMedium-reso%
lutionSpectra.III.MetallicityDistributionsofMilkyWayDwarfSatelliteGalaxies}
{Kirby}, E.~N., {Lanfranchi}, G.~A., {Simon}, J.~D., {Cohen}, J.~G., \&
  {Guhathakurta}, P. 2011, \apj, 727, 78

\bibitem[{{Kobulnicky} \&
  {Kewley}(2004)}]{Kobulnicky2004_Metallicitiesof0.3ltzlt1.0GalaxiesintheGOODS%
-NorthField}
{Kobulnicky}, H.~A. \& {Kewley}, L.~J. 2004, \apj, 617, 240

\bibitem[{{Larson}(1974)}]{Larson1974}
{Larson}, R.~B. 1974, \mnras, 169, 229

\bibitem[{{Laskar} {et~al.}(2011){Laskar}, {Berger}, \&
  {Chary}}]{Laskar2011_ExploringtheGalaxyMass-metallicityRelationatz3-5}
{Laskar}, T., {Berger}, E., \& {Chary}, R.-R. 2011, \apj, 739, 1

\bibitem[{{Leaman} {et~al.}(2013){Leaman}, {Venn}, {Brooks}, {Battaglia},
  {Cole}, {Ibata}, {Irwin}, {McConnachie}, {Mendel}, {Starkenburg}, \&
  {Tolstoy}}]{Leaman2013_TheComparativeChemicalEvolutionofanIsolatedDwarfGalax%
y:AVLTandKeckSpectroscopicSurveyofWLM}
{Leaman}, R., {Venn}, K.~A., {Brooks}, A.~M., {et~al.} 2013, \apj, 767, 131

\bibitem[{{Lee} {et~al.}(2006){Lee}, {Skillman}, {Cannon}, {Jackson}, {Gehrz},
  {Polomski}, \& {Woodward}}]{Lee2006}
{Lee}, H., {Skillman}, E.~D., {Cannon}, J.~M., {et~al.} 2006, \apj, 647, 970

\bibitem[{{Luck} {et~al.}(2011){Luck}, {Andrievsky}, {Kovtyukh}, {Gieren}, \&
  {Graczyk}}]{Luck2011_TheDistributionoftheElementsintheGalacticDisk.II.Azimut%
halandRadialVariationinAbundancesfromCepheids}
{Luck}, R.~E., {Andrievsky}, S.~M., {Kovtyukh}, V.~V., {Gieren}, W., \&
  {Graczyk}, D. 2011, \aj, 142, 51

\bibitem[{{Maiolino} {et~al.}(2008){Maiolino}, {Nagao}, {Grazian}, {Cocchia},
  {Marconi}, {Mannucci}, {Cimatti}, {Pipino}, {Ballero}, {Calura}, {Chiappini},
  {Fontana}, {Granato}, {Matteucci}, {Pastorini}, {Pentericci}, {Risaliti},
  {Salvati}, \& {Silva}}]{Maiolino2008}
{Maiolino}, R., {Nagao}, T., {Grazian}, A., {et~al.} 2008, \aap, 488, 463

\bibitem[{{Mannucci} {et~al.}(2010){Mannucci}, {Cresci}, {Maiolino}, {Marconi},
  \&
  {Gnerucci}}]{Mannucci2010_Afundamentalrelationbetweenmassstarformationratean%
dmetallicityinlocalandhigh-redshiftgalaxies}
{Mannucci}, F., {Cresci}, G., {Maiolino}, R., {Marconi}, A., \& {Gnerucci}, A.
  2010, \mnras, 408, 2115

\bibitem[{{McConnachie}(2012)}]{McConnachie2012_TheObservedPropertiesofDwarfGa%
laxiesinandaroundtheLocalGroup}
{McConnachie}, A.~W. 2012, \aj, 144, 4

\bibitem[{{Momany} {et~al.}(2005){Momany}, {Held}, {Saviane}, {Bedin},
  {Gullieuszik}, {Clemens}, {Rizzi}, {Rich}, \& {Kuijken}}]{Momany2005}
{Momany}, Y., {Held}, E.~V., {Saviane}, I., {et~al.} 2005, \aap, 439, 111

\bibitem[{{Nicholls} {et~al.}(2013){Nicholls}, {Dopita}, {Sutherland},
  {Kewley}, \&
  {Palay}}]{Nicholls2013_MeasuringNebularTemperatures:TheEffectofNewCollisionS%
trengthswithEquilibriumand$kappa$-distributedElectronEnergies}
{Nicholls}, D.~C., {Dopita}, M.~A., {Sutherland}, R.~S., {Kewley}, L.~J., \&
  {Palay}, E. 2013, \apjs, 207, 21

\bibitem[{{Pettini} \&
  {Pagel}(2004)}]{Pettini2004_[OIII]/[NII]asanabundanceindicatorathighredshift}
{Pettini}, M. \& {Pagel}, B.~E.~J. 2004, \mnras, 348, L59

\bibitem[{{Salzer} {et~al.}(2000){Salzer}, {Gronwall}, {Lipovetsky}, {Kniazev},
  {Moody}, {Boroson}, {Thuan}, {Izotov}, {Herrero}, \& {Frattare}}]{Salzer2000}
{Salzer}, J.~J., {Gronwall}, C., {Lipovetsky}, V.~A., {et~al.} 2000, \aj, 120,
  80

\bibitem[{{Savaglio} {et~al.}(2005){Savaglio}, {Glazebrook}, {Le Borgne},
  {Juneau}, {Abraham}, {Chen}, {Crampton}, {McCarthy}, {Carlberg}, {Marzke},
  {Roth}, {J{\o}rgensen}, \& {Murowinski}}]{Savaglio2005}
{Savaglio}, S., {Glazebrook}, K., {Le Borgne}, D., {et~al.} 2005, \apj, 635,
  260

\bibitem[{{Saviane} {et~al.}(2008){Saviane}, {Ivanov}, {Held}, {Alloin},
  {Rich}, {Bresolin}, \& {Rizzi}}]{Saviane2008}
{Saviane}, I., {Ivanov}, V.~D., {Held}, E.~V., {et~al.} 2008, \aap, 487, 901

\bibitem[{{Schmidt}(1959)}]{Schmidt1959_TheRateofStarFormation.}
{Schmidt}, M. 1959, \apj, 129, 243

\bibitem[{{Searle} \& {Sargent}(1972)}]{Searle1972}
{Searle}, L. \& {Sargent}, W.~L.~W. 1972, \apj, 173, 25

\bibitem[{{Tasca} {et~al.}(2009){Tasca}, {Kneib}, {Iovino}, {Le F{\`e}vre},
  {Kova{\v c}}, {Bolzonella}, {Lilly}, {Abraham}, {Cassata}, {Cucciati},
  {Guzzo}, {Tresse}, {Zamorani}, {Capak}, {Garilli}, {Scodeggio}, {Sheth},
  {Zucca}, {Carollo}, {Contini}, {Mainieri}, {Renzini}, {Bardelli},
  {Bongiorno}, {Caputi}, {Coppa}, {de La Torre}, {de Ravel}, {Franzetti},
  {Kampczyk}, {Knobel}, {Koekemoer}, {Lamareille}, {Le Borgne}, {Le Brun},
  {Maier}, {Mignoli}, {Pello}, {Peng}, {Perez Montero}, {Ricciardelli},
  {Silverman}, {Vergani}, {Tanaka}, {Abbas}, {Bottini}, {Cappi}, {Cimatti},
  {Ilbert}, {Leauthaud}, {Maccagni}, {Marinoni}, {McCracken}, {Memeo},
  {Meneux}, {Oesch}, {Porciani}, {Pozzetti}, {Scaramella}, \&
  {Scarlata}}]{Tasca2009_ThezCOSMOSredshiftsurvey:theroleofenvironmentandstell%
armassinshapingtheriseofthemorphology-densityrelationfromz1}
{Tasca}, L.~A.~M., {Kneib}, J.-P., {Iovino}, A., {et~al.} 2009, \aap, 503, 379

\bibitem[{{Tremonti} {et~al.}(2004){Tremonti}, {Heckman}, {Kauffmann},
  {Brinchmann}, {Charlot}, {White}, {Seibert}, {Peng}, {Schlegel}, {Uomoto},
  {Fukugita}, \& {Brinkmann}}]{Tremonti2004}
{Tremonti}, C.~A., {Heckman}, T.~M., {Kauffmann}, G., {et~al.} 2004, \apj, 613,
  898

\bibitem[{{Zahid} {et~al.}(2013){Zahid}, {Geller}, {Kewley}, {Hwang},
  {Fabricant}, \&
  {Kurtz}}]{Zahid2013_TheChemicalEvolutionofStar-formingGalaxiesovertheLast11B%
illionYears}
{Zahid}, H.~J., {Geller}, M.~J., {Kewley}, L.~J., {et~al.} 2013, \apjl, 771,
  L19

\end{thebibliography}

\end{document}